\documentclass[final,5p,twocolumn]{elsarticle}

\usepackage{graphicx}

\usepackage{amssymb}

\journal{Journal of Physics and Chemistry of Solids}
\newcommand{\Vec}[1]{\mbox{\boldmath$#1$}}
\begin{document}

\begin{frontmatter}



\title{Anion height as a controlling parameter for the superconductivity 
in iron pnictides and cuprates}


\author[1,2]{Kazuhiko Kuroki\corref{cor1}}\ead{kuroki@vivace.e-one.uec.ac.up}

\address[1]{Department of Applied Physics and Chemistry, 
The University of Electro-Communications, Chofu, Tokyo 182-8585, Japan}
\address[2]{JST, TRIP, Chofu, Tokyo 182-8585, Japan}

\cortext[cor1]{Corresponding author.}

\begin{abstract}
Both families of high $T_c$ superconductors, iron pnictides and 
cuprates, exhibit material dependence of superconductivity.
Here, we study its origin within the spin fluctuation pairing theory based 
on multiorbital models that take into account 
realistic band structures. For pnictides, we show that the 
presence and absence of Fermi surface pockets is sensitive to the 
pnictogen height measured from the iron plane due to the multiorbital 
nature of the system, which is reflected to the 
nodeless/nodal form of the superconducting gap and $T_c$. 
Surprisingly, even for the cuprates, which is conventionally modeled by 
a single orbital model, the multiorbital band structure is shown to 
play a crucial role in the material dependence of 
superconductivity. In fact, by 
adopting a two orbital model that considers the $d_{z^2}$ orbital 
on top of the $d_{x^2-y^2}$ orbital, we can resolve a long standing puzzle 
of why the single layered Hg cuprate have much higher $T_c$ than the 
La cuprate. Interestingly, here again the 
apical oxygen height measured from the CuO$_2$ plane 
plays an important role in determining the relative energy 
difference between $d_{x^2-y^2}$ and $d_{z^2}$ orbitals, 
thereby strongly affecting the superconductivity.
\end{abstract}


\end{frontmatter}



\section{Introduction}
Superconductivity in 
the iron-based pnictide LaFeAsO doped with fluorine 
discovered by Hosono's group\cite{Hosono,Ishidarev} 
and subsequent increase of the 
transition temperature ($T_c$) in the same family of compounds 
are remarkable as the first non-copper compound that 
has $T_c$'s exceeding 50 K.\cite{Ren}
Theoretical studies 
have shown that the electron-phonon coupling 
is too weak to account for the high $T_c$\cite{Boeri}, 
and exploring electronic pairing mechanisms has become a challenge. 
\begin{figure}[t]
\begin{center}
\includegraphics[width=6.5cm,clip]{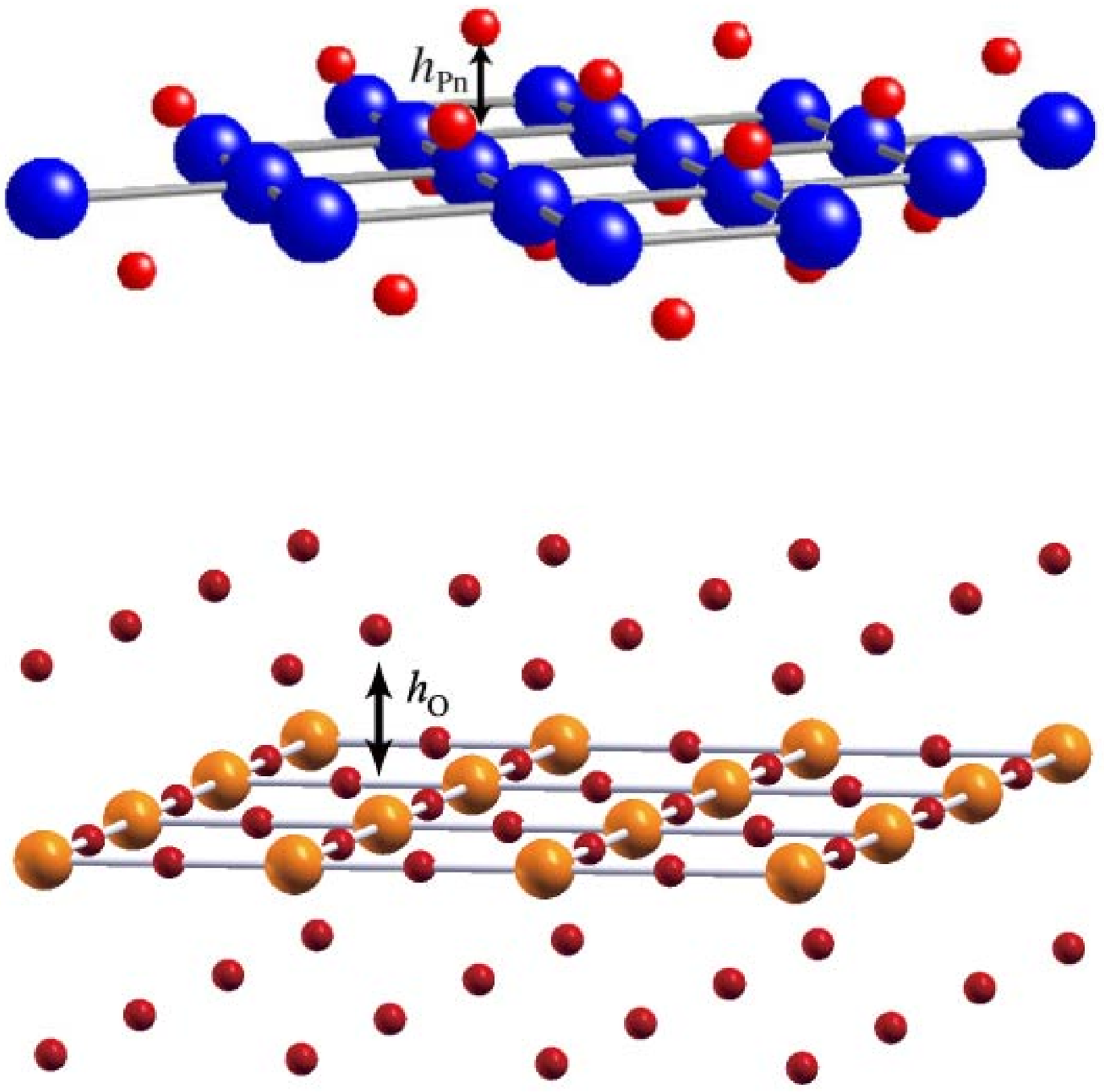}
\end{center}
{Fig.1 Anion height measured from the cation plane 
in the iron pnictides (top) and the cuprates (bottom).
\label{fig1}}
\end{figure}

An interesting observation of this series of material is their 
strong material dependence of superconductivity. 
$T_c$ ranges from 5K in LaFePO\cite{KamiharaP} to  55K\cite{Ren} in SmFeAsO, 
where the importance of the lattice structure has been pointed out\cite{Lee}.
A number of experiments on arsenides suggest fully-open superconducting gap,
consistent with the proposal of sign reversing $s$-wave pairing,
\cite{Mazin,1stpaper}
but experiments have shown presence of nodes in 
the superconducting gap of LaFePO\cite{Fletcher,Hicks,Matsuda}, 
suggesting that 
even the gap is material dependent. 

The strong material dependence of superconductivity in the 
pnictides reminds us of the high $T_c$ cuprates, which 
also exhibit strong material dependence of $T_c$.
It is well known that $T_c$ varies with the number of layers per unit cell, 
but an even more basic issue is the 
significant difference in $T_c$ within the {\it single-layered} materials,
i.e., La$_{2-x}$(Sr/Ba)$_x$CuO$_4$ with a maximum $T_c$ of about 40K 
versus HgBa$_2$CuO$_{4+\delta}$ with a $T_c \simeq 90$K. 
Phenomenologically, it has been recognized that the materials 
with $T_c\sim 100$K tend to have ``round'' Fermi surfaces,
while the Fermi surface of the ``low $T_c$'' 
La system is closer to a square shape which implies a relatively 
better nesting\cite{Pavarini,Tanaka}.

Conventionally, the cuprates with a rounded Fermi surface have been 
modeled by a {\it single-band} model with large second ($t_2(>0)$) and 
third ($t_3(<0)$) neighbor hopping integrals 
($(|t_2|+|t_3|)/|t_1|\sim 0.4$),  
while the La system has been considered to have 
smaller $t_2, t_3$ ($(|t_2|+|t_3|)/|t_1|\sim 0.1$). 
This, however, has brought about a contradiction between 
theories and experiments. 
Namely, while some phenomenological\cite{Moriya} 
and $t$-$J$ model\cite{Shih,Prelovsek} studies 
give a tendency consistent with the experiments, 
a number of many-body 
approaches for the {\it Hubbard-type} models with 
realistic values of on-site $U$ show suppression of superconductivity for 
large $t_2>0$ and/or $t_3<0$\cite{Scalapino,Maier,Kent}.

In this paper, we show that the experimentally 
observed material/lattice structure dependence of superconductivity 
in both pnictides\cite{2ndpaper} and cuprates\cite{Sakakibara} 
can be understood 
by analyzing the superconductivity 
within the spin fluctuation pairing theory that takes into account the 
realistic band structure. 
A surprising finding is that not only the pnictides, but also 
the cuprates have band structures where {\it multiple} $d$ orbitals play 
an important role. Interestingly, in both families, 
the anion height measured from the cation plane (Fig.1) turns out to be 
one of the key parameters 
that controls the multiorbital band structure and thus the 
superconductivity. These studies show that a combination of 
effective model construction based on first principles band calculation 
and the application of many body theory can give a 
realistic description on the 
material dependence of superconductivity in correlated systems.

\section{Construction of multiorbital models}
\subsection{Pnictides}
We start with the band structure of 
LaFeAsO. LaFeAsO has a layered structure, where Fe atoms form a 
square lattice in each layer, which is 
sandwiched by As atoms.
Due to the tetrahedral coordination of As atoms, 
there are two Fe atoms per unit cell.  
The experimentally determined lattice constants 
are $a=4.03$\AA \ and $c=8.74$\AA, with 
two internal coordinates $z_{\rm La}=0.142$ and $z_{\rm As}=0.651$.\cite{Cruz}
We have first obtained the band structure 
with plane-wave basis\cite{pwscf},
which is then used to 
construct the maximally localized 
Wannier functions\cite{MaxLoc}. 
These Wannier functions 
have five orbital symmetries ($d_{3Z^2-R^2}$, 
$d_{XZ}$, $d_{YZ}$, $d_{X^2-Y^2}$, 
$d_{XY}$, where $X, Y, Z$ refer to those for this unit cell 
with two Fe sites as shown in Fig.2(a). 
The two Wannier orbitals in 
each unit cell are equivalent in that each Fe atom has the same 
local arrangement of other atoms.
We can then take a unit cell that 
contains only one orbital per symmetry by 
unfolding the Brillouin zone,
and we end up with an effective five-band model on a 
square lattice, where 
$x$ and $y$ axes are rotated by 
45 degrees from $X$-$Y$.  
We refer all the wave vectors in the unfolded Brillouin 
zone hereafter. 
We define the band filling $n$ as the number of electrons/number of sites
(e.g., $n=10$ for full filling). 
The doping level $x$ 
in LaFeAsO$_{1-x}$F$_x$ is related to the band filling as $n=6+x$.

\begin{figure}[t]
\begin{center}
\includegraphics[width=6.0cm,clip]{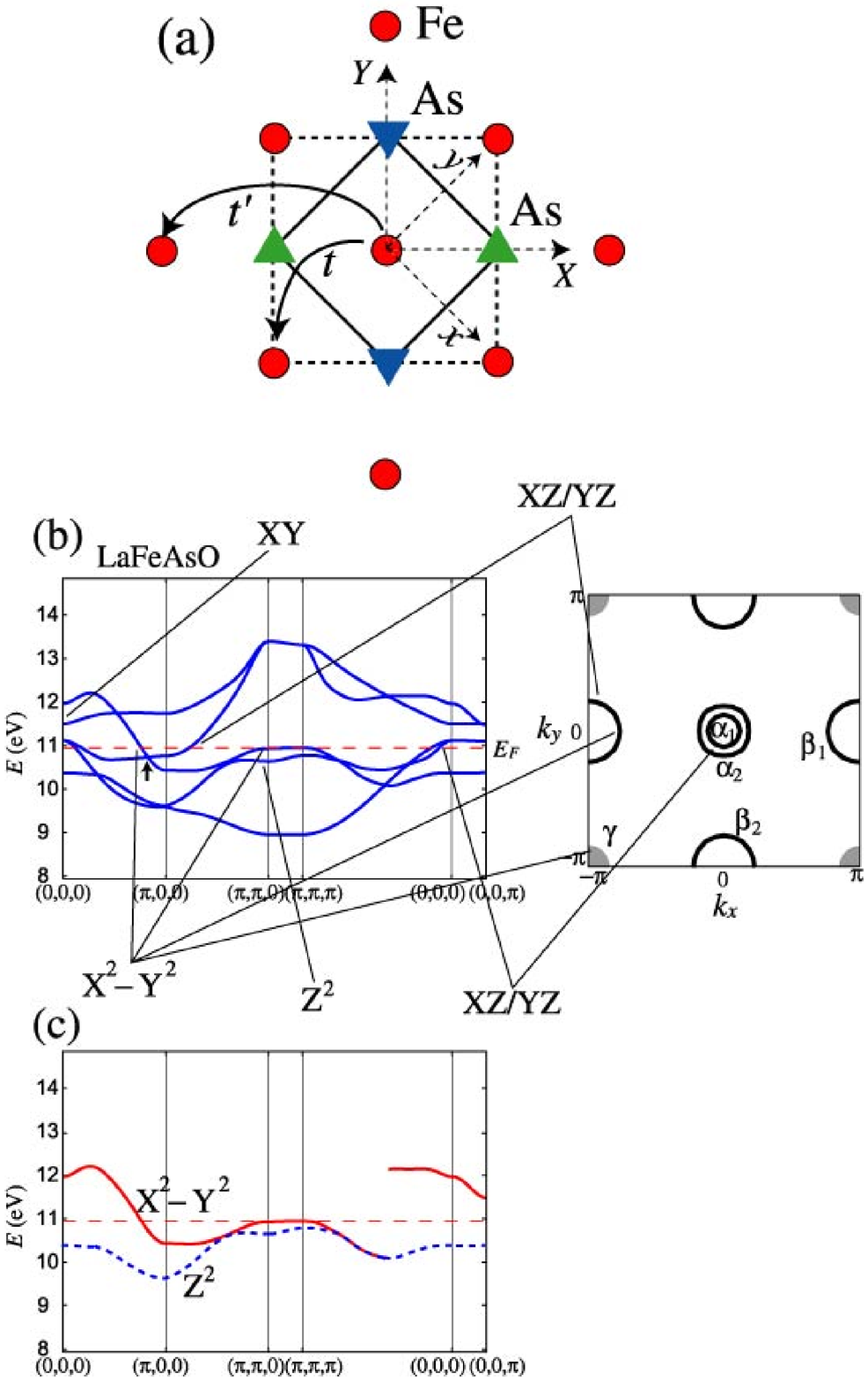}
\end{center}
{Fig.2 (a) The original (dashed lines) and 
reduced (solid) unit cells with $\bullet$ (Fe), $\nabla$ (As 
below the Fe plane) and $\triangle$ (above Fe). 
(b) The band structure (left) of the five-band model for LaFeAsO, and the 
Fermi surface (right) at $k_z=0$ for $n=6.1$. The main orbital 
characters of some portions of the bands and the Fermi surface 
are indicated.  
The dashed horizontal line in the band structure indicates the Fermi level 
for $n=6.1$. The short arrow in the band structure indicates the 
position of the Dirac cone closest to the Fermi level.  
The gray areas in the Fermi surface 
around the zone corners represent the $\gamma$ Fermi surface.  
(c) The portion of the band that has mainly the $d_{X^2-Y^2}$ (solid red) 
and $d_{Z^2}$ (dashed blue) orbital character.
\label{fig2}}
\end{figure}

The five bands are heavily 
entangled as shown in Fig.2(b) (left)
reflecting strong hybridization 
of the five $3d$ orbitals, which suggests that the 
minimal model for the pnictides should include all the five orbitals. 
In Fig.2(b) (right), 
the Fermi surface for $n=6.1$ (corresponding to $x=0.1$) 
obtained by ignoring the inter-layer hoppings 
is shown in the two-dimensional unfolded Brillouin zone.
The Fermi surface consists of four pieces:   
two concentric hole pockets (denoted here as $\alpha_1$, $\alpha_2$) 
centered around $(k_x, k_y)=(0,0)$, two electron pockets 
around $(\pi,0)$ $(\beta_1)$ or $(0,\pi)$ $(\beta_2)$, respectively. 
$\alpha_i$ ($\beta_i$) corresponds to the 
Fermi surface around the $\Gamma$Z line (MA in the original Brillouin zone) 
in the first-principles band calculation.\cite{Singh}
Besides these pieces of the Fermi surface, there is a portion of the band 
near $(\pi,\pi)$ that 
touches the $E_F$, so that 
the portion acts as a ``quasi Fermi surface $(\gamma)$'' around $(\pi,\pi)$, 
which has in fact an important contribution to the spin susceptibility 
as we shall see below. 
As for the orbital character, $\alpha$ and portions of $\beta$ near 
Brillouin zone edge have mainly $d_{XZ}$ and $d_{YZ}$ character, 
while the portions of 
$\beta$ away from the Brillouin zone edge and $\gamma$ have 
mainly $d_{X^2-Y^2}$ orbital character (see also Fig.2(c)).

To be more precise, the above band structure is sensitive to the 
lattice structure.
In the upper two panels of Fig.3, 
we compare the band structure of LaFePO and NdFeAsO. 
A large difference between the two materials lies in the band structure near 
the wave vector $(\pi,\pi)$. In LaFePO, 
the $\gamma$ Fermi surface originating from the 
$d_{X^2-Y^2}$ band around $(\pi,\pi)$ largely 
sinks below the Fermi level, 
and in turn the $d_{Z^2}$ band, which is not effective for superconductivity,
rises up. On the other hand, in 
NdFeAsO, the $d_{X^2-Y^2}$ band rises up even more than in LaFeAsO.
The origin of this band structure variation is mainly due to the 
variation of the pnictogen height $h_{\rm Pn}$
\cite{2ndpaper,Singh,GeorgesArita,Lebegue},
which varies from $h_{\rm Pn}=1.14 {\rm \AA}$ in LaFePO to 
$h_{\rm Pn}=1.38{\rm \AA}$ in NdFeAsO.\cite{Lee}
In fact, in the lower two panels of Fig.3, we show band structures of 
LaFeAsO with hypothetical lattice structures, where we fix the 
lattice constants and change only the height to those of NdFeAsO or LaFePO.
We can see that the band structure around $(\pi,\pi)$, namely, 
the position of the $d_{X^2-Y^2}$ and $d_{Z^2}$ bands are 
determined by the pnictogen height. 

The origin of this height  
sensitivity of the band structure 
is the following. In Fig.2(c), we plot the $d_{X^2-Y^2}$ and 
$d_{Z^2}$ portions of the bands. Around $(\pi,\pi)$, the lower portion of 
$d_{X^2-Y^2}$ and the upper portion of $d_{Z^2}$ lies close to the $E_F$. 
When the pnictogen height becomes low, the Wannier orbitals widely spread 
toward the pnictogen, so that the band width tends to become large.
Therefore, the $d_{X^2-Y^2}$ portion around $(\pi,\pi)$ sinks below 
$E_F$, while the $d_{Z^2}$ portion rises up.

\begin{figure}[t]
\begin{center}
\includegraphics[width=6.0cm,clip]{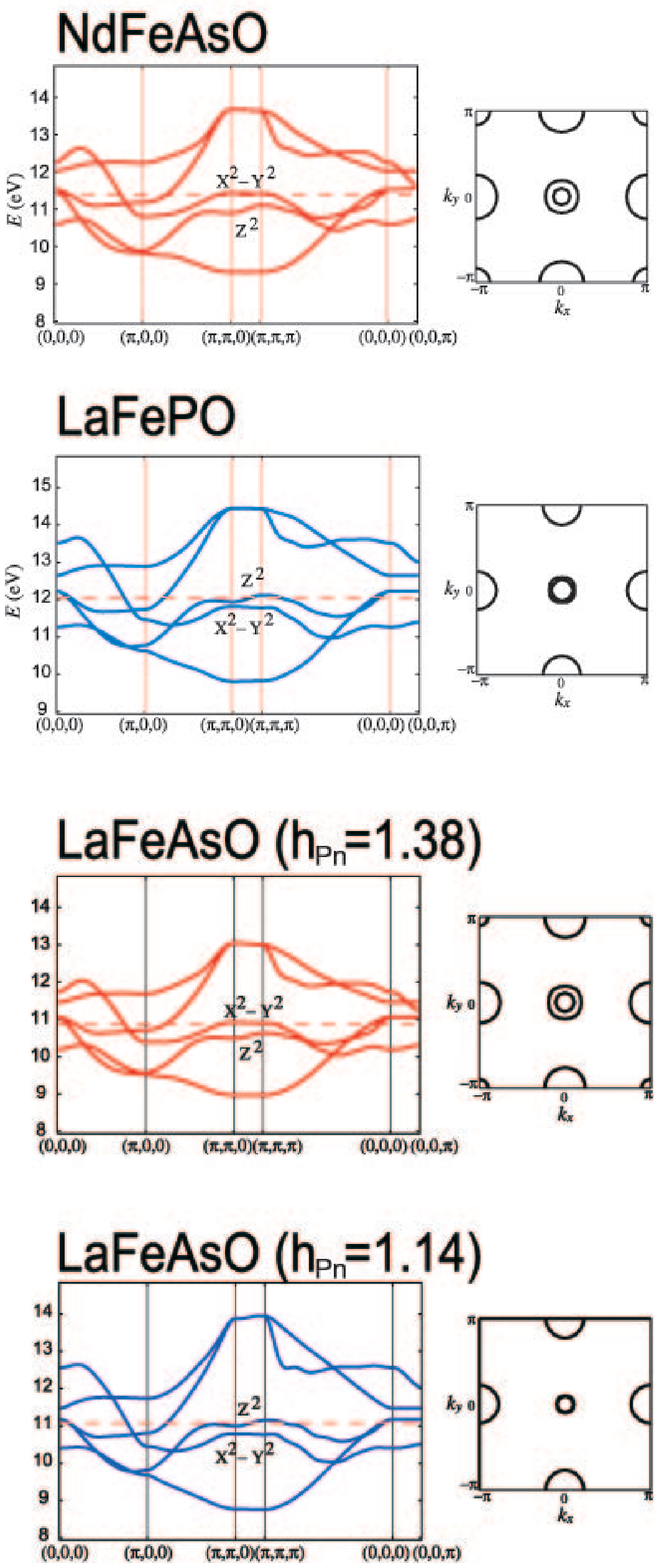}
\end{center}
{Fig.3  From top to bottom :The band structure and the Fermi surfaces of 
NdFeAsO, LaFePO, hypothetical structure of LaFeAsO with 
$h_{Pn}=1.38 {\rm \AA}$ and $1.14 {\rm \AA}$. 
\label{fig3}}
\end{figure}

\subsection{Cuprates}
In contrast to the pnictides, theoretical studies on cuprates 
has mostly been done for single band models, or the three band 
model that takes into account the oxygen orbitals that strongly 
hybridizes with the Cu $d_{x^2-y^2}$ orbital. As mentioned in the 
introduction,  within these single 
orbital approaches, it is difficult to understand 
the experimental observation that the more 
Fermi surface is rounded, higher the $T_c$.
To resolve the discrepancy between experimental observations and theory,
we consider a {\it two-orbital} 
model, where we take into account the $d_{z^2}$ 
orbital on top of the $d_{x^2-y^2}$\cite{Sakakibara}.   
In fact, for the La system, it has long been 
known that a band with a strong $d_{z^2}$ 
character lies rather close to the Fermi energy\cite{Shiraishi,Eto,Freeman}.
More recently, it has been discussed in refs.\cite{Andersen,Pavarini} 
that the shape of the Fermi surface is determined by the energy level of the 
``axial state'' consisting of a mixture of Cu $d_{z^2}$-O $p_z$ and 
Cu $4s$ orbitals, and 
that the strength of the $d_{z^2}$ contribution 
causes the difference in the 
Fermi surface shape between the La and Hg systems. 
Namely, the $d_{z^2}$ contribution is large in the 
La system making the Fermi surface closer to a square, 
while the contribution is small in the Hg system making the Fermi surface 
more rounded.   In Fig.4, we show the 
first-principles\cite{pwscf} result for 
band structures in the two-orbital model for the La and Hg systems, 
obtained by constructing maximally localized Wannier 
orbitals\cite{MaxLoc}.  
The lattice parameters adopted here are experimentally determined 
ones for the doped materials\cite{La-st,Hg-st}.
We can confirm that in the La system 
the main band (usually considered 
as the ``$d_{x^2-y^2}$ band'') has in fact a strong $d_{z^2}$ character 
on the Fermi surface near the N point, which corresponds to the wave vectors 
 $(\pi,0), (0,\pi)$ in the Brillouin zone of the square lattice.  
The $d_{z^2}$ contribution is seen to ``push up'' 
 the van Hove singularity (vHS) of the main band, resulting in a 
seemingly well nested (square shaped) 
Fermi surface.  In the Hg system, on the other hand, the $d_{z^2}$ band 
stays well away from $E_F$, and consequently the vHS 
is lowered, resulting in  a rounded Fermi surface.
\begin{figure}[!b]
\begin{center}
\includegraphics[width=7cm]{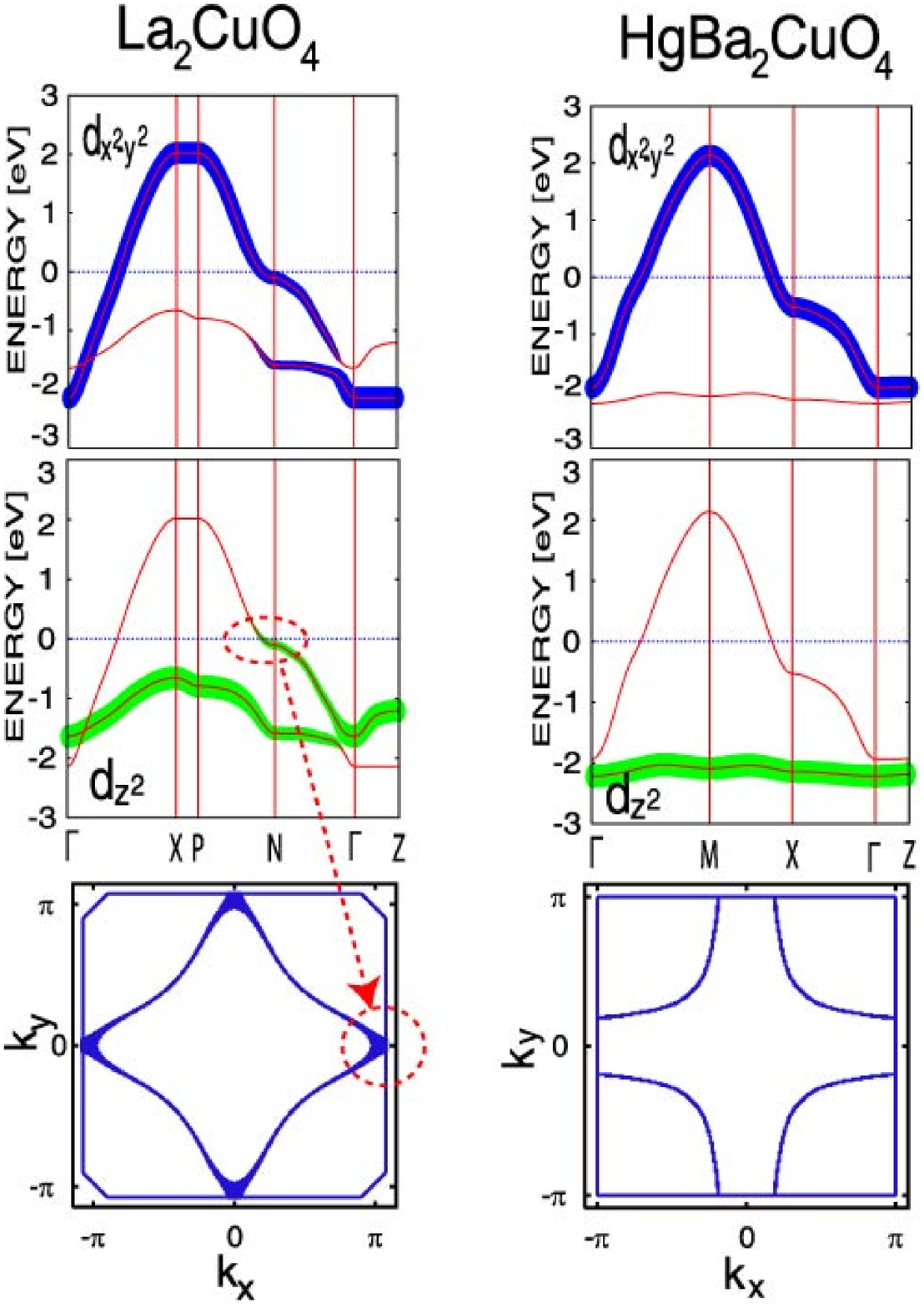}
\end{center}
{Fig.4 The band structure of the two 
($d_{x^2-y^2}$-$d_{z^2}$) orbital model for La$_2$CuO$_4$ (left) and 
HgBa$_2$CuO$_4$ (right). The top (middle) panels 
depict the strength of the $d_{x^2-y^2}$ 
($d_{z^2}$) characters with thickened lines, while the bottom panels the 
 Fermi surfaces (for a total band filling $n=2.85$).
}
\label{fig4}
\end{figure}

If we estimate in the two-orbital model   
the ratio $(|t_2|+|t_3|)/|t_1|$ 
 {\it within the $d_{x^2-y^2}$ orbitals}, we get 
0.35 for the La system against 
0.41 for Hg, which are rather close to each other. 
This  contrasts with the situation in which the model 
is constrained into a single band model that considers 
one kind of Wannier orbital to reproduce the main 
band that intersects the Fermi energy.   
There, the Wannier orbital has mainly $d_{x^2-y^2}$ character,
but has ``tails'' with a $d_{z^2}$ character especially for the La system. 
Then the ratio $(|t_2|+|t_3|)/|t_1|$ in the single-orbital model 
reduces to 0.14 for La against 0.37 for Hg, 
which is just the conventional view 
mentioned in the Introduction.  
From this, we can confirm that  it is the $d_{z^2}$ 
contribution that makes the Fermi 
surface in the La system square shaped, while the 
``intrinsic'' Fermi surface of the 
high $T_c$ cuprate family is, as in the Hg system, 
rounded (which is actually 
due to the Cu $4s$ orbital as shown in Ref.\cite{Andersen,Pavarini},
whose contribution is effectively considered in the present model).

\section{Many-body Hamiltonian}

For the many body part of the Hamiltonian, 
we consider the standard interaction terms that comprise 
the intra-orbital Coulomb $U$, the inter-orbital 
Coulomb $U'$, the Hund's coupling $J$ and the pair-hopping $J'$. 
The many body Hamiltonian then reads
\begin{eqnarray}
H &=& \sum_i\sum_\mu\sum_{\sigma}\varepsilon_\mu n_{i\mu\sigma}
 + 
\sum_{ij}\sum_{\mu\nu}\sum_{\sigma}t_{ij}^{\mu\nu}
c_{i\mu\sigma}^{\dagger} c_{j\nu\sigma}\nonumber\\
&+&\sum_i\left[U\sum_\mu n_{i\mu\uparrow} n_{i\mu\downarrow}
+U'\sum_{\mu > \nu}\sum_{\sigma,\sigma'} n_{i\mu\sigma} n_{i\mu\sigma'}
\right.\nonumber\\
&-&\left.J\sum_{\mu\neq\nu}\Vec{S}_{i\mu}\cdot\Vec{S}_{i\nu}
+J'\sum_{\mu\neq\nu} c_{i\mu\uparrow}^\dagger c_{i\mu\downarrow}^\dagger
c_{i\nu\downarrow}c_{i\nu\uparrow}
\right], 
\end{eqnarray}
where $i,j$ denote the sites and $\mu,\nu$ the orbitals, 
and $t_{ij}^{\mu\nu}$ is the transfer energy obtained from the 
maximally localized Wannier orbitals. 

For these models, we apply spin fluctuation theory and 
solve the Eliashberg equation to analyze superconductivity.
Multiorbital random phase approximation (RPA) is described in e.g.
Refs.\cite{Yada,Takimoto}. In the present case, 
Green's function $G_{lm}(k)$ $(k=(\Vec{k},i\omega_n))$
is a $m\times m$ matrix, where $m$ is the number of orbitals. 
The irreducible susceptibility matrix 
\begin{equation} 
\chi^0_{l_1,l_2,l_3,l_4}(q) =-\sum_k G_{l_1l_3}(k+q)G_{l_4l_2}(k)
\end{equation}
$(l_i = 1,...,m)$ has $m^2\times m^2$ components, and 
the spin and the charge (orbital) susceptibility matrices are obtained 
from matrix equations, 
\begin{equation}
\chi_s(q)=\frac{\chi^0(q)}{1-S\chi^0(q)}
\end{equation}
\begin{equation}
\chi_c(q)=\frac{\chi^0(q)}{1+C\chi^0(q)}
\end{equation}
where 
\[
S_{l_1l_2,l_3l_4},\;\; C_{l_1l_2,l_3l_4}
=\left\{\begin{array}{ccc}
U,& U &\;\; l_1=l_2=l_3=l_4\\ 
U',&-U'+J & \;\; l_1=l_3\neq l_2=l_4\\
J,&2U'-J,&\;\; l_1=l_2\neq l_3=l_4\\
J',&J'& \;\; l_1=l_4\neq l_2=l_3\end{array}  \right.
\]
The Green's function and 
the effective singlet pairing interaction, 
\begin{equation}
V^s(q)=\frac{3}{2}S\chi^s(q)S-\frac{1}{2}C\chi^c(q)C+\frac{1}{2}(S+C),
\end{equation}
are plugged into the linearized Eliashberg equation, 
\begin{eqnarray}
\lambda\phi_{l_1l_4}(k)&=&-\frac{T}{N}\sum_q
\sum_{l_2l_3l_5l_6}V_{l_1 l_2 l_3 l_4}(q)\times\nonumber\\
&&G_{l_2l_5}(k-q)\phi_{l_5l_6}(k-q)
G_{l_3l_6}(q-k).
\end{eqnarray}
Here, $T_c$ is the temperature where the eigenvalue $\lambda$ reaches 
unity, so when $\lambda$ is calculated at a fixed temperature for 
different situations/materials, the value can be considered 
as a qualitative measure for the $T_c$.

\section{Anion height dependence of superconductivity}
\subsection{Pnictides}
We first look into the superconductivity of pnictides.
We apply RPA to the 5 orbital model, where we take 
$U=1.2$, $U'=0.9$, $J=J'=0.15$eV, 
the temperature is fixed at $T=0.02$\ eV  and 
$32\times32\times 4$ $k$-point meshes and 512 
Matsubara frequencies are taken. 
Let us first look at the the spin susceptibility of LaFeAsO. 
In the top panels of Fig.5,  
we show $\chi_{s3333}$ and $\chi_{s4444}$,  
which represent the spin correlation 
within $d_{YZ}$ and $d_{X^2-Y^2}$ orbitals, respectively.
$\chi_{s3333}$ peaks solely around $(\pi,0)$ and $(0,\pi)$, 
which reflects the nesting between the 
$XZ,YZ$-charactered portions of 
$\alpha$ and $\beta$ Fermi pockets as shown in a bottom panel 
of Fig.5. On the other hand, $\chi_{s4444}$ has peaks around $(\pi,0),(0,\pi)$ 
and around $(\pi,\pi/2)/(\pi/2,\pi)$ as well. 
The former is due to the nesting 
between the $\gamma$ quasi Fermi surface and the $d_{X^2-Y^2}$ 
portion of the $\beta$ Fermi surface as was first pointed out in 
ref.\cite{Mazin},  while the latter originates 
from the nesting between the $d_{X^2-Y^2}$ 
portion of the $\beta_1$ and $\beta_2$ Fermi surfaces.\cite{1stpaper}
The $(\pi,0),(0,\pi)$ feature is 
consistent with the stripe (i.e., collinear) antiferromagnetic order for the 
undoped case, which was suggested by transport 
and optical reflectance\cite{Dong},  
and further confirmed by neutron scattering experiments.\cite{Cruz}
The stabilization of such an antiferromagnetic ordering has also been 
pointed out in first principles calculations.\cite{Ishibashi,Dong,Mazin2}
Furthermore, the presence of 
spin fluctuations near the wavevector  $(\pi,0),(0,\pi)$  
{\it in the unfolded Brillouin zone }
has indeed been confirmed in an inelastic neutron scattering 
experiment\cite{Lumsden}.

\begin{figure}[t]
\begin{center}
\includegraphics[width=7.0cm,clip]{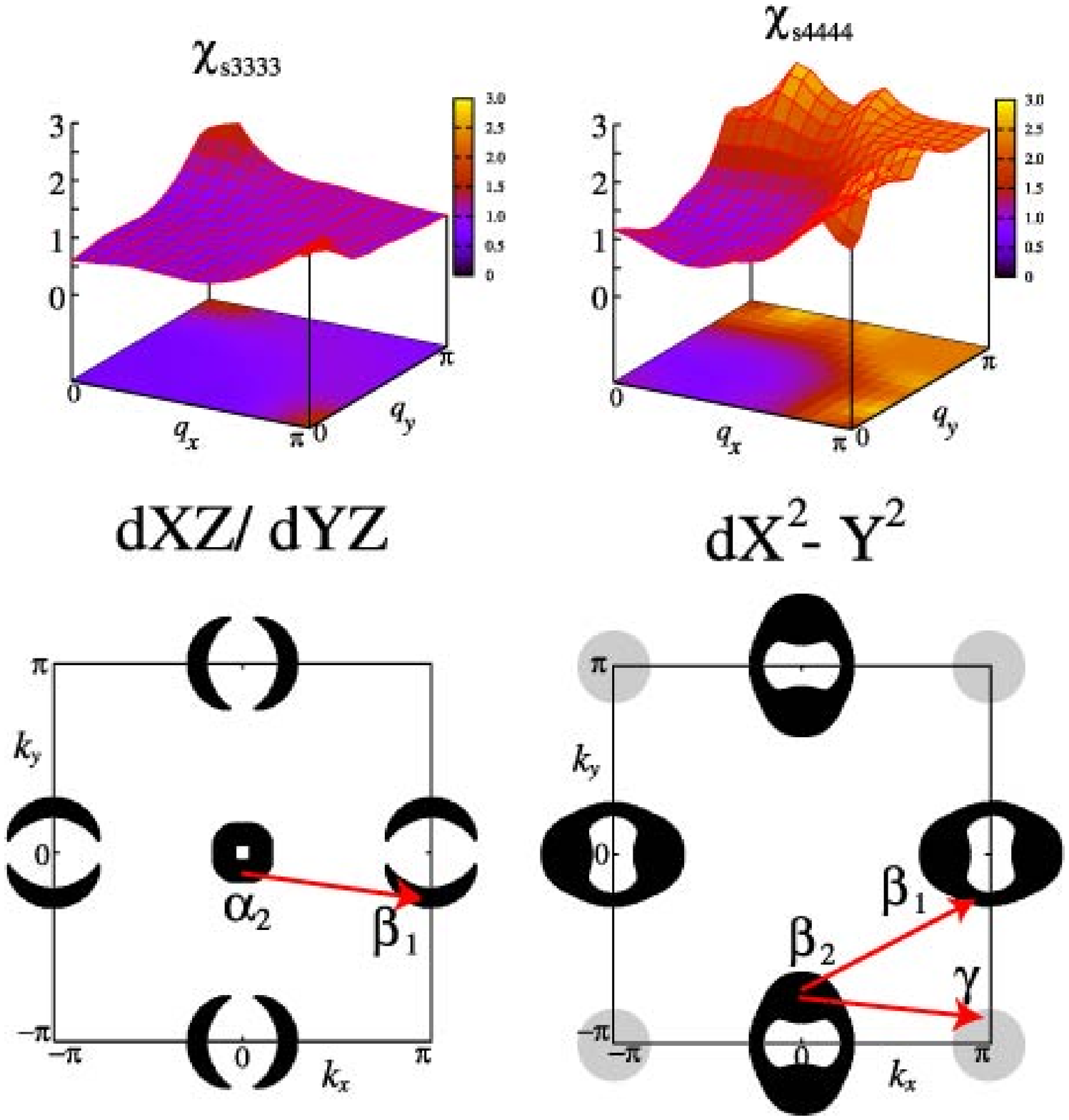}
\end{center}
{Fig.5 Top panel: RPA result 
of the diagonal component of the spin susceptibility matrix, 
 $\chi_{s3333}$(left) and $\chi_{s4444}$(right)
($3: YZ, 4: X^2-Y^2$) for LaFeAsO.  
Bottom panel: Nesting is shown for the Fermi surface 
for orbitals $XZ, YZ$ (left)
and for $X^2-Y^2$ (right). Here the thickness of the Fermi surface 
represents the strength of the respective orbital character. The gray areas 
around the corners in the left panel indicates the 
$\gamma$ ``quasi'' Fermi surface. 
\label{fig5}}
\end{figure}

When the pairing is mediated by spin fluctuations, 
the gap on the Fermi surface tends to change its sign across 
the wave vector at which the spin fluctuations develop.
Since there are multiple spin fluctuation modes in our model, 
the superconducting gap 
should be determined by the cooperation or competition between these multiple 
modes.
Specifically, the $\alpha$-$\beta$ and $\gamma$-$\beta$ 
nestings tend to favor the fully-gapped, sign-reversing $s$-wave 
($s\pm$-wave), in 
which the gap changes sign between $\alpha$ and $\beta$ and also 
between $\gamma$ and $\beta$, but has 
a constant sign on each pocket as shown in 
Fig.5.\cite{Mazin}.
This kind of  gap has in fact been considered in the context of 
raising $T_c$ in spin fluctuation mediated pairing in that 
the nodes of the gap, which is a necessary evil for 
spin fluctuation mediated pairing, do not have to intersect the 
Fermi surface\cite{KA,KA2}. 
On the other hand, $\beta_1$-$\beta_2$ nesting tends to change the 
sign of the gap between these pockets, which can result in 
either $d$-wave or an $s$-wave pairing with nodes on the 
$\beta$ Fermi surface, as shown in Fig.5
\cite{Graser,2ndpaper}.
\begin{figure}[t]
\begin{center}
\includegraphics[width=7.0cm,clip]{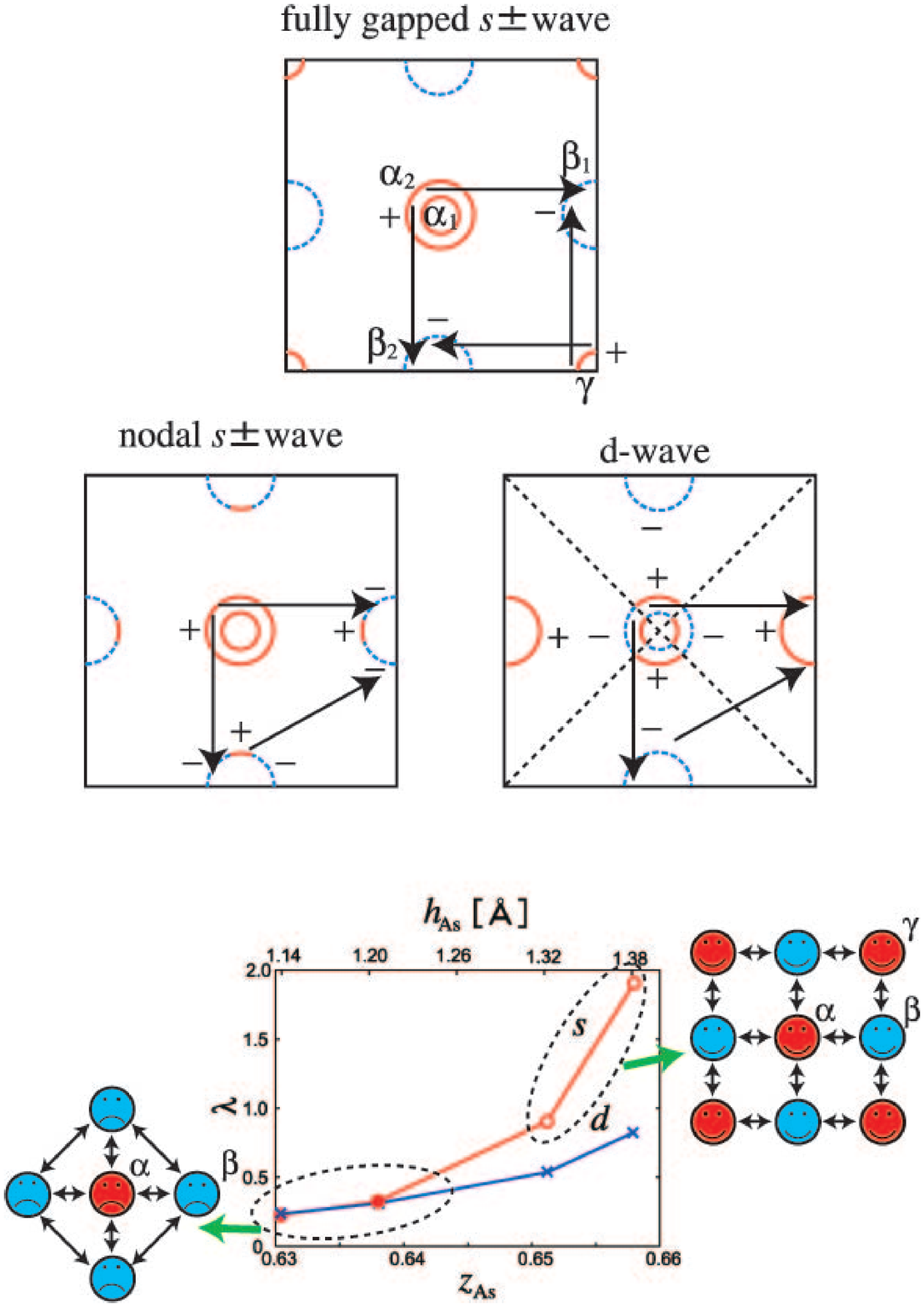}
\end{center}
{Fig.6 Top and middle panels: the fully-gapped $s\pm$ wave, the nodal 
$s\pm$ wave and $d$ wave gap 
are schematically shown. The solid red (dashed blue) curves 
represent positive (negative) sign of the gap. The arrows 
indicate the dominating nesting vectors. Bottom: the $s$-wave and 
$d$-wave eigenvalues of the 
Eliashberg equation plotted as functions of As height for the 
hypothetical lattice structure of LaFeAsO. The figures in the 
left and right insets show schematically the presence or absence of 
frustration.
\label{fig6}}
\end{figure}

When the pnictogen atom is at low positions,
the $\beta-\gamma$ nesting is not effective, 
and the competition between $(\pi/0)/(0,\pi)$ and $(\pi,\pi/2)/(\pi/2,\pi)$ 
spin fluctuations results in a frustration in momentum space, where the gap 
tends to change the sign of the gap across these vectors. 
Consequently, the $s$-wave gap has nodes intersecting the $\beta$ Fermi 
surface, and this pairing is nearly degenerate with $d$-wave pairing.
When the pnictogen is high, the $\beta$-$\gamma$ 
nesting becomes more effective than the $\beta_1$-$\beta_2$ nesting, 
and the two kinds of of $(\pi/0)/(0,\pi)$ spin fluctuations,
those originating from $d_{XZ}/d_{YZ}$ and $d_{X^2-Y^2}$ orbitals,  
cooperates without experiencing 
frustration to result in the fully gapped $s\pm$-wave pairing.

The absence/presence of 
the frustration affects also the strength of the 
superconducting instability, where the fully gapped 
case has a stronger tendency toward pairing. This can be seen from the 
eigenvalue of the Eliashberg equation plotted as a function of the 
pnictogen height in the bottom panel of Fig.6,
where we adopt the 5 band model for LaFeAsO, but 
vary the height.  Thus the pnictogen height 
$h_{\rm Pn}$ acts as a switch 
between higher $T_c$ nodeless and low $T_c$ nodal pairings, 
which explains the difference between 
NdFeAsO and LaFePO in both $T_c$  and the gap. 

Here, we have focused on the effect of the $\gamma$ Fermi surface 
around $(\pi,\pi)$ in the context of the lattice structure dependence, 
but its effect has recently been studied in detail from the viewpoint 
of the electron-electron interaction and the band filling
\cite{Kemper,Ikeda3}. Also, a real space picture for the 
presence/absence of the nodes in the gap has been discussed\cite{Kariyado}.
The electron correlation effects beyond RPA has also been taken 
into account in the fluctuation exchange (FLEX) study\cite{Ikeda3,Ikeda2} 
and a functional renormalization group study\cite{DHLee}, which 
also finds a similar effect of the 
$\gamma$ Fermi surface on the presence/absence of nodes in the 
superconducting gap.

The above scenario is consistent with the 
presence of line nodes in the superconducting 
gap of the low $T_c$ phosphides
\cite{Fletcher,Hicks,Matsuda}.
On the other hand, 
the fully gapped $s\pm$-wave is difficult to detect experimentally because the 
absolute value of the gap 
is essentially the same as that of a simple $s$-wave gap, with no 
characteristic angular dependence. Nevertheless, a recent 
phase sensitive STM experiment has clearly detected the 
sign reverse between the Fermi surfaces separated by 
$(\pi,0),(0,\pi)$\cite{Hanaguri}.

\subsection{Cuprates}
We now move on to the superconductivity in the cuprates.
For the electron-electron interactions, we take $U=3.0$eV,  
 $J=J'=0.3$eV, which gives the interorbital $U'=U-2J=2.4$eV.
The temperature is fixed at $k_BT=0.01$eV.
As for the band filling (number of electrons/site), 
we concentrate on the total 
$n=2.85$, for which the main band has 0.85.
Here we apply the multiorbital FLEX\cite{Yada,Bickers,Dahm}, 
which takes into account the self-energy correction to the 
Green's function self-consistently, 
for the three-dimensional lattice 
taking $32\times 32\times 4$ $k$-point meshes and 1024 Matsubara frequencies.

Here again, we focus on how the anion height, namely, the 
apical oxygen height measured from the CuO$_2$ plane affects the band 
structure and hence superconductivity. 
This is motivated by the fact that the 
energy level offset between $d_{x^2-y^2}$ and $d_{z^2}$ orbitals  
should be controlled (at least partially)
by the ligand field, hence by the height, $h_{\rm O}$, of the 
apical oxygen\cite{Eto}.  
To single out this effect, 
let us examine the two-orbital model of the La cuprate for which 
we increase $h_{\rm O}$ from its 
original value $2.41$\AA\hspace{0.1cm}with other lattice parameters fixed.  
In Fig.7(a), we plot the Eliashberg equation eigenvalue of the 
$d$-wave superconductivity  
as a function of $h_{\rm O}$.  We can see that 
$\lambda$ monotonically increases with the height.
As seen from the inset of Fig.7(b), 
$\Delta E$, defined as the difference of the on-site energy 
between $d_{x^2-y^2}$ and $d_{z^2}$ orbitals, is positively correlated with 
$h_{\rm O}$ as expected, and importantly, Fig.7(b) 
shows that the increase in $\lambda$ 
is positively correlated with the increase in $\Delta E$.
Hence the superconductivity turns out to be enhanced 
as the  $d_{z^2}$ band moves away from the main band.  
Note that this occurs {\it despite the 
Fermi surface becoming more rounded} with larger $\Delta E$, 
namely, the effect of the 
orbital character (smaller $d_{z^2}$ contribution) 
dominates over the Fermi surface shape effect.  
Conversely, the strong $d_{z^2}$ orbital character 
in the Fermi surface around the wave vectors  
$(\pi,0),(0,\pi)$ works destructively against $d$-wave 
superconductivity. Physically, this may be understood as 
follows. $d$-wave superconductivity occurs due to pair scattering 
from $\sim (\pi,0)$ to $\sim (0,\pi)$ and vice versa 
mediated by antiferromagnetic spin fluctuations (Fig.7(c)). When the 
states around $(\pi,0),(0,\pi)$ has strong $d_{z^2}$ character, 
superconductivity is suppressed since 
$d$-wave pairing 
has a  rough tendency for higher $T_c$ in 
bands that are nearly half filled, whereas the 
$d_{z^2}$ orbital here is nearly full filled 
(has holes only around $(\pi,0), (0,\pi)$).

In Fig.7, we have also plotted the 
corresponding values of the Hg system  
obtained with the actual lattice structure. 
We can see that $\Delta E$ in Hg is indeed larger than that in La as 
expected, but actually 
$\Delta E\simeq 2.2$eV for Hg is 
larger than $\Delta E\simeq 1.3$eV, which is the value the La system 
would take for $h_{\rm O}=2.8{\rm \AA}$.
Consequently, $\lambda$ 
for Hg is somewhat larger than 
that for the La system with the same value of $h_{\rm O}$.  
This implies that there are also some effects other than the apical oxygen 
height that also enhance $\Delta E$ (i.e., lower the $d_{z^2}$ level with 
respect to the $d_{x^2-y^2}$ level) 
in the Hg system, thereby further favoring 
$d$-wave superconductivity.  In this context, 
the present result reminds us of the so-called 
``Maekawa's plot'', where a positive correlation between $T_c$ 
and the level of the apical oxygen $p_z$ hole was 
pointed out\cite{Maekawa}.  
Since a higher $p_z$ hole level (i.e., a lower 
$p_z$ electron level) is likely to 
lower $E_{z^2}$ because the $d_{z^2}$ Wannier orbital in our model 
consists mainly of Cu$3d_{z^2}$ and apical oxygen $2p_z$ orbitals, 
the positive correlation between 
$\Delta E$ and $T_c$ found here is indeed 
consistent with Maekawa's plot.  
\begin{figure}[t]
\begin{center}
\includegraphics[width=6.5cm]{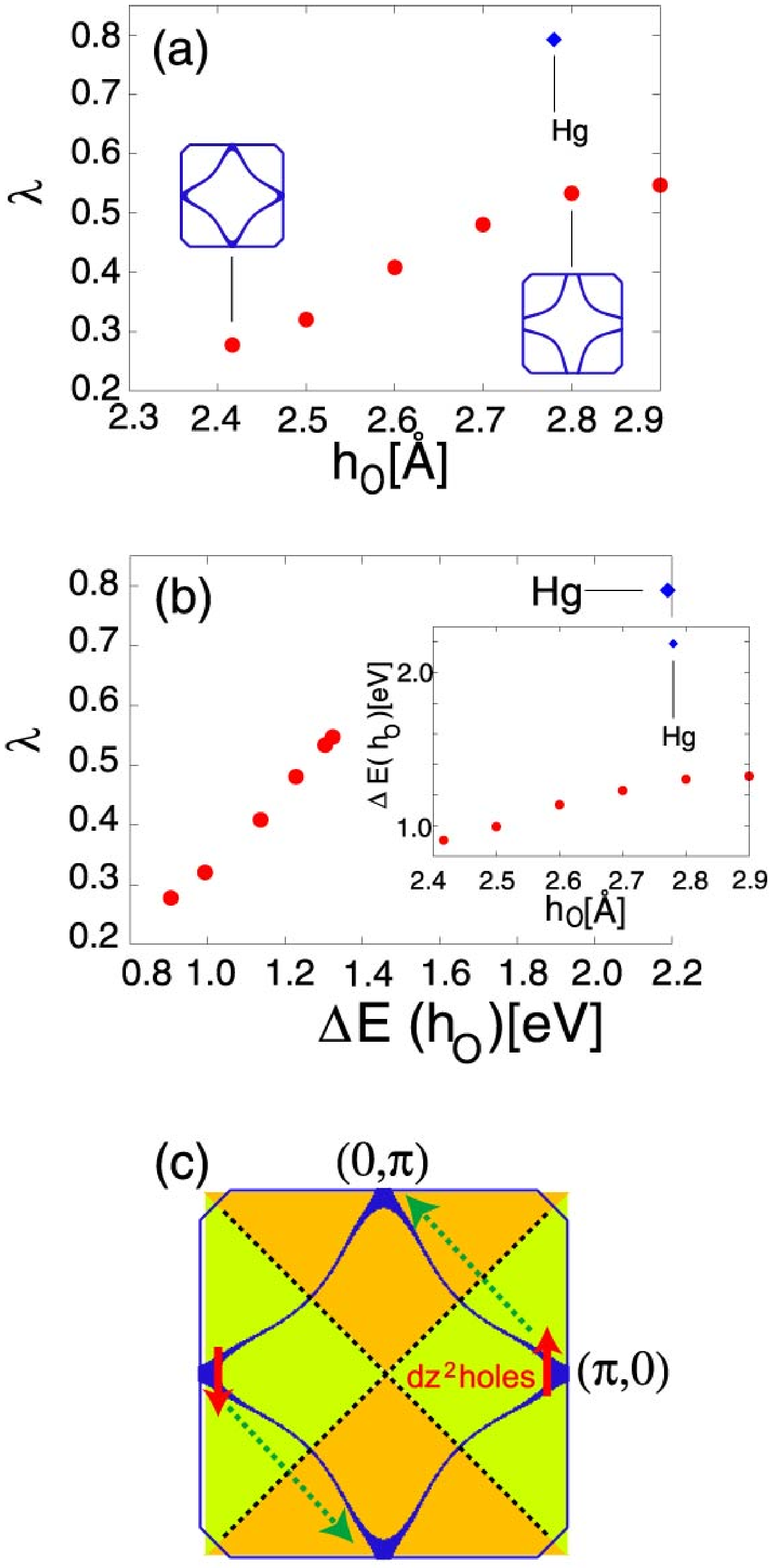}
\end{center}
{Fig.7 The eigenvalue of the Eliashberg equation 
$\lambda$ (red circles) when $h_{\rm O}$(a) or $\Delta E$(b) is varied 
hypothetically in the lattice structure of La$_2$CuO$_4$. 
Blue diamond indicates the eigenvalue of HgBa$_2$CuO$_4$.
Inset in (b) shows the relation  between $h_{\rm O}$ and $\Delta E$. 
In (c), the pair scattering of $d_{z^2}$ holes around $(\pi,0)$ 
$(0,\pi)$ is illustrated. 
}
\label{fig7}
\end{figure}

\section{Conclusion}
In this paper, we have focused on the origin of the 
material dependence of superconductivity in the two families of 
high $T_c$ superconductors, pnictides and cuprates.
We have analyzed the superconductivity within the 
spin fluctuation pairing theory applied to  models that 
take into account the realistic multiorbital band structure.
It is shown that not only in the iron pnictides, but even in the 
cuprates, the multiorbital 
band structure is affected by the lattice structure, 
which in turn affects superconductivity.
It is interesting to note that in both families, the 
anion height measured from the cation plane plays an important role, 
although its effect on the band structure appears in a different 
manner : in the cuprates the the energy difference between 
$d_{x^2-y^2}$ and $d_{z^2}$ is affected by the ligand field, 
while in the pnictides, the relative position between the 
$d_{X^2-Y^2}$ and $d_{Z^2}$ bands are affected mainly due to the 
variation of the band width, namely, the former (latter) band 
lies in the upper (lower) portion of the band structure, 
so that when the band width becomes wider upon lowering the 
height, the location of the 
lower band edge of $d_{X^2-Y^2}$ and that of the upper band 
edge of $d_{Z^2}$ (both around the wave vector $(\pi,\pi)$ )is exchanged.
The good description of the material dependence of the superconductivity 
within the present approach suggests commonality between the 
two families in that the spin degrees of freedom are playing an important role 
in the occurrence of superconductivity\cite{Scalapino2}, 
although there are differences in the strength of the electron 
correlation.

The present studies show that a combination of 
effective model construction based on first principles band calculation 
and application of many body theory can give a realistic description on the 
material dependence of superconductivity in correlated systems, 
and may even suggest a possibility of 
theoretically predicting new superconductors in the near future.
\ \\

\noindent
{\bf Acknowledgment}

The author would like to thank Hideo Aoki, Ryotaro Arita, and 
Hidetomo Usui for collaboration. The author also wishes to 
thank Seiichiro Onari and Katsuhiro Suzuki 
for collaboration in the study of pnictides, and 
Hirofumi Sakakibara in the study of cuprates.
Numerical calculations were performed at 
the Information Technology Center, University of Tokyo, 
and at the Supercomputer Center,
ISSP, University of Tokyo. 
This study has been supported by 
Grants-in-Aid for Scientific Research from  MEXT of Japan and from 
the Japan Society for the Promotion of Science.




\begin{thebibliography}{00}
\bibitem{Hosono}
Y.Kamihara, T.Watanabe, M.Hirano, H. Hosono: 
J. Am. Chem. Soc. {\bf 130} (2008) 3296.
\bibitem{Ishidarev} for a review, see e.g., K. Ishida, Y. Nakai, 
and H. Hosono : J. Phys. Soc. Jpn. {\bf 78}, 062001 (2009).
\bibitem{Ren} Z.-A Ren, W. Lu, J. Yang, W. Yi, X.-L. Shen, Z.-C. Li, 
G.-C. Che, X.-L Dong, L.-L. Sun, F. Zhou, Z.-X. Zhao: 
Chin. Phys. Lett. {\bf 25} (2008) 2215.
\bibitem{Boeri} L. Boeri {\it et al.}, Phys. Rev. Lett. {\bf 101}, 
026403  (2008).
\bibitem{KamiharaP} Y. Kamihara {\it et al.}, 
J. Am. Chem. Soc. {\bf 128}, 10012 (2006).
{\it ibid} {\bf 102}, 109902(E) (2009). 
\bibitem{Lee} C.-H. Lee {\it et al.} J. Phys. Soc. Jpn.{\bf 77}, 083704 (2008).
\bibitem{Mazin} I.I. Mazin, D.J. Singh, M.D. Johannes, M.H. Du:: 
Phys. Rev. Lett. {\bf 101} (2008) 057003.
\bibitem{1stpaper} K. Kuroki, S. Onari, R. Arita, H. Usui, Y. Tanaka, 
H. Kontani, and H. Aoki : 
Phys. Rev. Lett. {\bf 101} (2008) 087004, 
\bibitem{Fletcher} J.D. Fletcher, A. Serafin, L. Malone, J. Analytis, 
J-H Chu, A.S. Erickson, I.R. Fisher, and 
A. Carrington, Phys. Rev. Lett. {\bf 102}, 147001 (2009).
\bibitem{Hicks} C.W.Hicks, T.M. Lippman, M.E. Huber, J.G. Analytis, J.-H. Chu, 
A.S. Erickson, I.R.Fisher, and K.A.Moler, arXiv: 0903.5260.
\bibitem{Matsuda} M. Yamashita, N. Nakata, Y. Senshu, S. Tonegawa, 
K. Ikada, K. Hashimoto, H. Sugawara, T. Shibauchi, and Y. Matsuda :  
arXiv: 0906.0622.
\bibitem{Pavarini}
E. Pavarini {\it et al.}: Phys. Rev. Lett. {\bf 87}, 047003 (2001).
\bibitem{Tanaka} K. Tanaka {\it et al.}: Phys. Rev. B {\bf 70}, 092503 (2004).
\bibitem{Moriya} T. Moriya and K. Ueda: J. Phys. Soc. Jpn. {\bf 63}, 1871 
(1994).
\bibitem{Shih} 
C.T. Shih {\it et al.}: Phys. Rev. Lett. {\bf 92}, 227002 (2004). 
\bibitem{Prelovsek} P. Prelov$\rm \check{s}$ek and A. Ram$\rm \check{s}$ak, 
Phys. Rev. B {\bf 72}, 012510 (2005).
\bibitem{Scalapino} For a review, see D.J. Scalapino :  
{\it Handbook of High Temperature Superconductivity}, Chapter 13, 
Eds. J.R. Schrieffer and J.S. Brooks (Springer, New York, 2007).
\bibitem{Maier} Th. Maier {\it et al.}: Phys. Rev. Lett. {\bf 85}, 1524 (2000).
\bibitem{Kent} P.R.C. Kent {\it et al.}: Phys. Rev. B {\bf 78}, 035132 (2008).
\bibitem{2ndpaper} K. Kuroki, H. Usui, S. Onari, R. Arita: 
and H. Aoki: Phys. Rev. B {\bf 79}, 224511 (2009).
\bibitem{Sakakibara}
H. Sakakibara, H. Usui, K. Kuroki, R. Arita, and H. Aoki: Phys. Rev. Lett. 
{\bf 105}, 057003 (2010).
\bibitem{Cruz} C.de la Cruz, Q. Huang, J. W. Lynn, Jiying Li, 
W. Ratcliff II, J. L. Zarestky, H. A. Mook, G. F. Chen, J. L. Luo, 
N. L. Wang, P. Dai: Nature {\bf 453} (2008) 899.
\bibitem{MaxLoc} N. Marzari and D. Vanderbilt: Phys. Rev. B 
{\bf 56} (1997) 12847; 
I. Souza, N. Marzari and D. Vanderbilt: 
Phys. Rev. B {\bf 65} (2002) 035109.
The Wannier functions are generated by the code developed by
A. A. Mostofi, J. R. Yates, N. Marzari, I. Souza and D. Vanderbilt:
(http://www.wannier.org/). 
\bibitem{pwscf} 
S. Baroni, 
A. Dal Corso, S. de Gironcoli, P. Giannozzi,
C. Cavazzoni, G. Ballabio, S. Scandolo, G. Chiarotti, P. Focher,
A. Pasquarello, K. Laasonen, A. Trave, R. Car, N. Marzari,
A. Kokalj: http://www.pwscf.org/.
\bibitem{Singh} D.J. Singh and M.-H. Du: Phys. Rev. Lett. {\bf 100} (2008) 
237003. 
\bibitem{GeorgesArita} V. Vildosola, L. Pourovskii, R. Arita, S. Biermann, and 
A. Georges: Phys. Rev. B {\bf 78}, 064518 (2008).
\bibitem{Lebegue} S. Lebegue, Z.P. Yin, and W.E. Pickett: New J. Phys. 
{\bf 11}, 025004 (2009).
\bibitem{Shiraishi} K. Shiraishi {\it et al.}: Solid State Commun. {\bf 66},
629 (1988).
\bibitem{Eto} H. Kamimura and M. Eto: J. Phys. Soc. Jpn. {\bf 59}, 3053 
(1990); M. Eto and H. Kamimura: J. Phys. Soc. Jpn. {\bf 60}, 2311 (1991).
\bibitem{Freeman} A.J. Freeman and J. Yu: Physica B {\bf 150}, 50 (1988).
\bibitem{Andersen} O.K. Andersen {\it et al.}: J. Phys. Chem. Solids 
{\bf 56}, 1573 (1995).
\bibitem{La-st} 
J.D. Jorgensen {\it et al.}: Phys. Rev. Lett. {\bf 58}, 1024 (1987).
\bibitem{Hg-st}
J.L. Wagner {\it et al.}: Physica C {\bf 210}, 447 (1993).
\bibitem{Yada} K. Yada and H. Kontani: J. Phys. Soc. Jpn. {\bf 74}
(2005) 2161.
\bibitem{Takimoto} T. Takimoto, T. Hotta, and K. Ueda: 
Phys. Rev. B {\bf 69} (2004) 104504.
\bibitem{Dong} J. Dong, H.J. Zhang, G. Xu, Z. Li, W.Z. Hu, D. Wu, G.F.Chen, 
X. Dai, J.L. Luo, Z.Fang, and N.L.Wang:  
Europhys. Lett. {\bf 83}  (2008) 27006.
\bibitem{Ishibashi} S. Ishibashi, K. Terakura, H. Hosono: 
J. Phys. Soc. Jpn. {\bf 77} (2008) 053709.
\bibitem{Mazin2}
I.I. Mazin, M.D. Johannes, L. Boeri, K. Koepernik, 
D.J. Singh: Phys. Rev. B {\bf 78}  (2008) 085104.
\bibitem{Lumsden} M.D.Lumsden, A.D.Christianson, E.A.Goremychkin, S.E. Nagler,
H.A.Mook, M.B.Stone, D.L.Abemathy, T.Guidi,G.J.MacDougali, C. dela Cruz, 
A.S. Sefat, M.A. McGuire, B.C.Sales, and D. Mandrus: Nat. Phys. {\bf 6}, 182 
(2010).
\bibitem{KA} K. Kuroki and R. Arita: Phys. Rev. B {\bf 64} (2001) 024501.
\bibitem{KA2} K. Kuroki, T. Kimura, and R. Arita: 
Phys. Rev. B {\bf 66} (2002) 184508.
\bibitem{Graser} S.Graser, T.A. Maier, P.J. Hirshfeld, and D.J. Scalapino :  
New J. Phys. {\bf 11}, 025016 (2009).
\bibitem{Kemper} A.F.Kemper, T.A.Maier, S.Graser, H.-P. Cheng, 
P.J. Hirshfeld, and D.J. Scalapino: arXiv:1003.2777.
\bibitem{Ikeda3} H. Ikeda, R. Arita, and J. Kunes, arXiv: 1002.4471.
\bibitem{Kariyado} T. Kariyado and M. Ogata: J. Phys. Soc. Jpn. {\bf 79}, 
0337803 (2010).
\bibitem{Ikeda2} H. Ikeda, R. Arita, and J. Kunes: Phys. Rev. B {\bf 81}, 
054502 (2010).
\bibitem{DHLee} F. Wang, H. Zhai, D.-H. Lee: Phys. Rev. B {\bf 81}, 184512 
(2010).
\bibitem{Hanaguri} T. Hanaguri, S. Niitaka, K. Kuroki, and H. Takagi: 
Science {\bf 328}, 474 (2010).
\bibitem{Bickers} N.E. Bickers, D.J. Scalapino, and S.R. White :  
Phys. Rev. Lett. {\bf 62}, 961 (1989).
\bibitem{Dahm} T. Dahm and L. Tewordt: Phys. Rev. Lett. {\bf 74}, 793 (1995)
\bibitem{Maekawa} Y. Ohta {\it et al.},  
Phys. Rev. B {\bf 43}, 2968 (1991).
\bibitem{Scalapino2} D.J. Scalapino: axXiv:1002.2413.
\end{thebibliography}
\end{document}